\documentclass[aps,prb,twocolumn,letterpaper]{revtex4-1}

\usepackage{graphicx}
\usepackage{amsmath}
\usepackage{amsfonts,amssymb}
\usepackage{graphicx,graphics,csquotes,epstopdf}
\usepackage{color,textcomp,bm,subfigure,array,xspace,siunitx}

\usepackage[pdfstartview=FitV,bookmarks=true,
            linktocpage=true,colorlinks=true,
            urlcolor=blue,linkcolor=red,citecolor=blue]{hyperref}

\newcommand{\ie}{\textit{i.e.}\@\xspace}

\newcommand{\nm}{\si{\nano\metre}\@\xspace}
\newcommand{\um}{\si{\micro\metre}\@\xspace}
\newcommand{\dbus}{\si{\decibel/\micro\second}\@\xspace}
\newcommand{\umm}{\si{\micro\meter^2}\@\xspace}
\newcommand{\mm}{\si{\milli\metre}\@\xspace}

\newcommand{\MWcm}{\si{\mega\watt/\centi\square\metre}\@\xspace}

\newcommand{\ms}{\si{\milli\second}\@\xspace}
\newcommand{\ns}{\si{\nano\second}\@\xspace}
\newcommand{\Hz}{\si{\hertz}\@\xspace}

\newcommand{\etal}{\emph{et al.~}}

\begin{document}

\title{Laser beam shaping for enhanced Zero-Group Velocity Lamb modes generation}

\author{Fran\c{c}ois Bruno}     	
\author{J\'er\^ome Laurent}     	
\author{Paul Jehanno}    		 
\author{Daniel Royer}       		
\author{Claire Prada}       		\email{claire.prada@espci.fr}
\affiliation{Institut Langevin, ESPCI Paris, CNRS (UMR 7587), PSL Research University, Paris, France}

\begin{abstract}
Optimization of Lamb modes induced by laser can be achieved by adjusting the spatial source distribution to the mode wavelength ($\lambda$). The excitability of Zero-Group Velocity (ZGV) resonances in isotropic plates is investigated both theoretically and experimentally for axially symmetric sources. Optimal parameters and amplitude gains are derived analytically for spot and annular sources of either Gaussian or rectangular energy profiles. For a Gaussian spot source, the optimal radius is found to be $\lambda_{ZGV}/\pi$. Annular sources increase the amplitude by at least a factor of $3$ compared to the optimal Gaussian source. Rectangular energy profiles provide higher gain than Gaussian ones. These predictions are confirmed by semi-analytical simulation of the thermoelastic generation of Lamb waves, including the effect of material attenuation. Experimentally, Gaussian ring sources of controlled width and radius are produced with an axicon-lens system. Measured optimal geometric parameters obtained for Gaussian and annular beams are in good agreement with theoretical predictions. A ZGV resonance amplification factor of $2.1$ is obtained with the Gaussian ring. Such source should facilitate the inspection of highly attenuating plates made of low ablation threshold materials like composites.
\end{abstract}

\date{15 september 2016}
\maketitle

\section{Introduction}\label{sec:intro}

Ultrasonic techniques are widespread for the characterization of mechanical properties of materials. Guided Lamb modes are often used to determine the elastic properties of plate-like structures. It is well known that for some Lamb modes, the group velocity vanishes at finite wavelength.\cite{tassoulas1984wave} It was demonstrated that these specific Zero-Group Velocity (ZGV) modes are very well generated and detected in metallic plates by laser-based ultrasonic techniques~\cite{Prada05a,Clorennec06,clorennec07}. The energy deposited by either continuous laser source~\cite{Prada05a} or pulsed laser impact~\cite{Clorennec06} remains trapped under the source, resulting in local and narrow ZGV resonances. The frequencies of these resonances depend on plate thickness and bulk acoustic wave velocities in such a way that broadband and local measurements of ZGV resonances in an isotropic material provide Poisson's ratio.\cite{clorennec07} ZGV modes also exist in multi-layered plates where they are useful to characterize the bonding between the layers.\citep{cho14,mezil14,mezil15} For these various applications, it is important to optimize the laser source geometry in order to increase the ZGV mode amplitude and to obtain high signal-to-noise ratio. In general, the Lamb mode amplitude increases with the total amount of deposited energy. However, for most applications, ultrasonic generation has to remain non-destructive. This constraint limits the deposited energy density to the ablation threshold, which is, for example, about $10$~\MWcm in Duralumin.
 
Different solutions have been considered to overcome this limitation. Several studies proposed to used multiple laser sources synchronized with appropriate time delays to mimic phased array systems and to focus bulk waves.\citep{noroy1993laser,steckenrider1995sensitivity} However, these systems are cumbersome and expensive. With a single thermoelastic source, it is necessary to shape the laser beam in order to generate a particular guided mode.

In the past, surface acoustic waves (SAWs) have been efficiently generated by using laser interference patterns~\cite{cachier1970optical} or by splitting a laser beam into regularly spaced line sources with an optical diffraction grating.\cite{huang1992laser} Recently, Grunsteildl~\etal used an intensity modulated laser combined with spatial light modulator (SLM) to adjust the excitation both spatially and temporally. With the laser intensity being spread on parallel, equidistant lines, this technique allows an efficient and selective generation of a specific mode with low peak power densities on the sample.\cite{grunsteidl13}

A solution to enhance SAW amplitude at a particular point is to use an annular beam. This was achieved by Cielo~\etal with an axicon-lens system.\cite{CieloNadeauLamontagne1985_Ultrasonics} A pulsed laser beam was focused to a sharp ring on an aluminium sample to generate a convergent Rayleigh wave. The wave amplitude at the center of convergence was increased by a factor of 20 with respect to a line-source produced wave with equal surface heating. Focused bulk ultrasonic waves generated by ring-shaped laser beam were also applied to flaw detection.\cite{WangLittman1996_JAP} This solution was explored on thick samples and the measurement of Rayleigh and bulk acoustic waves were compared to numerical results obtained by thermoelastic finite element modelling.\cite{DixonPetcher2012_JAP} Recently, ring laser sources were also obtained with a SLM.\cite{zamiri2014converging}

Optimization of the laser source to enhance ZGV Lamb modes was first discussed by Balogun~\etal in a study of the generation of the $S_1S_2$-ZGV resonance by an amplitude modulated laser source.\cite{BalogunMurrayPrada2007_JAP} Using a semi-analytical model,\cite{SpicerMcKieWagner1990_APL} the $S_1S_2$-ZGV resonance amplitude was calculated as a function of the radius $\mathrm{w}$ of the Gaussian beam: $f(r)=\exp(-r^2/\mathrm{w}^2)$. They showed numerically that, for an aluminum plate, the optimal radius is about 1.3 times the plate thickness and about $0.3$ times the mode's wavelength $\lambda_{S_1S_2}$. This question was then addressed by Grunsteidl~\etal experimentally and numerically using a time domain finite difference technique.\cite{Grunsteidl15} They found, for tungsten, an optimal radius between $1.0$ and $1.1$ times the plate thickness.\\

In the present work, the generations of ZGV modes using different axi-symmetric beam profiles are studied. A theoretical analysis is proposed to find the optimal parameters of different source shapes. Then, experiments achieved for Gaussian and annular sources are presented. Quasi-Gaussian annular beams of controlled width and diameter are shaped with an axicon-lens system. Unlike previous studies using thin rings of radii that are large compared to SAW wavelength,\citep{CieloNadeauLamontagne1985_Ultrasonics,WangLittman1996_JAP} thick rings with radii on the order of ZGV mode wavelength are produced. In Sec.~\ref{sec1}, the model of thermoelastic generation in a plate with an axi-symmetric source is recalled. In Sec.~\ref{sec2}, the theoretical approach allows us to establish an analytical formulation of the optimal width and radii versus the mode wavelength for a Gaussian source, an annular beam of Gaussian profile, a top-hat beam, and an annular beam of square profile. The semi-analytical simulation developed by Balogun \etal \cite{BalogunMurrayPrada2007_JAP} was adapted to the different source shapes and used to calculate the displacement amplitudes at ZGV resonance frequencies as a function of the source geometric parameters. Numerical results are then compared with the analytical ones. Finally, in Sec.~\ref{sec3}, experimental measurements achieved on a $1$-\mm thick Duralumin plate are described and compared with theoretical results. The generation of $S_1S_2$-ZGV Lamb mode is optimized by adjusting the axicon-lens system.  

\section{Lamb mode generation with an axi-symmetric thermoelastic source}\label{sec1}

In order to calculate the amplitude of ZGV resonances excited by an axi-symmetric laser source, we use the semi-analytical model introduced in Ref.~\cite{SpicerMcKieWagner1990_APL} to describe thermoelastic conversion and the coupling with Lamb modes as in Balogun~\etal.\cite{BalogunMurrayPrada2007_JAP} It is assumed that the temperature field is independent of the elastic field. Thus, the heat equation is solved and the temperature field is considered as a source term in the elastodynamic equations. Assuming that the optical penetration depth $\gamma$ is smaller than the plate thickness $2h$, the absorbed power density $P_a$ for a laser source at the surface $z=-h$ can be written in cylindrical coordinates $(r,z)$ 
\begin{equation}
P_a(r,z,t) = E_{tot}f(r)\left( \frac{e^{-\frac{z+h}{\gamma}}}{\gamma}\right )g(t), 
\label{eq_AbsorbedPowerDensity}
\end{equation}
where $E_{tot}$ is the total energy absorbed by the plate, $f(r)$ is the spatial distribution of the laser intensity normalized to unity: $2\pi\bigl\lmoustache _{0}^\infty f(r)rdr =1$ and $g(t)$ is the normalized laser pulse profile: $\bigl\lmoustache _{0}^\infty g(t)dt =1$. Then the spatial energy density distribution on the plate surface is $E(r)= E_{tot}f(r)$. In practice, in order to avoid abblation phenomena the maximum surface energy density, equal to
 \begin{equation}
 I= E_{tot} \max(f(r)),
  \label{eq_EnergyMax}
 \end{equation}
 is limited by the abblation threshold of the material $I_a$. The temperature rise is linked to the absorbed power density through the heat equation: 
\begin{equation}
\nabla^2 T - \frac{1}{\kappa}\frac{\partial T}{\partial t}=-\frac{P_a}{K}, \notag
\label{eq_ThermalEquation}
\end{equation}
where $K$ is the thermal conductivity, $\kappa = K / \rho C$ the thermal diffusivity, $\rho$ is the material density, and $C$ is the specific heat. As the problem is axi-symmetric, it can be solved using the Fourier-Hankel transforms of the power density $\overline{P_a}^{H_0}(k,z,\omega)$ and the temperature rise $\overline{T}^{H_0}(k,z,\omega)$ where $k$ is the wave number and $\omega$ is the angular frequency. This transformation leads to the following differential equation: 
\begin{equation}
\frac{\partial^2 \overline{T}^{H_0} }{\partial z^2} - \chi^2 \overline{T}^{H_0} = -\frac{\overline{P}_a^{H_0}}{K},
{\rm \quad with \quad }  \chi^2 = k^2 + i \frac{\omega}{\kappa},\\ 
\label{eq_HankelTemperatureEq}
\end{equation}
where  
\begin{equation}
\overline{P}_a^{H_0}(k,z,\omega) = \left(\frac{e^{-\frac{z+h}{\gamma}}}{\gamma}\right) E^{H_0}(k) \overline{g}(\omega).
\label{eq_HankelAbsorbedPowerDensity}
\end{equation}
Considering that in our experiments, the laser pulse is very short (a few nanoseconds) compared with the period of the studied Lamb modes, we omit the function $\overline{g}(\omega)$ in the following. The term $E^{H_0}(k)\exp(-h/\gamma)$ is factorized in Eq.~\eqref{eq_HankelTemperatureEq} so that the solution can be expressed as 
\begin{equation}
\small{
\overline{T}_{H_0} = \left[ T_1e^{\chi z} + T_2e^{- \chi z} - \left(\frac{\gamma}{1-\gamma^2  \chi^2}\right)e^{-z/\gamma}\right] E^{H_0}(k)\frac{e^{-h/\gamma}}{K}}.
\label{eq_HankelTemperature}
\end{equation}
The constants $T_1$ and $T_2$ are determined from boundary conditions. Neglecting heat diffusion into the air, the absence of any thermal flux on each plate surface $z = \pm h$ implies 
\begin{equation}
{\left . \frac{\partial T}{\partial z}  \right |}_{z = \pm h} = 0 
\Leftrightarrow 
{\left . \frac{\partial \overline{T}^{H_0}}{\partial z}  \right |}_{z = \pm h} = 0.
\label{eq_BCThermalEquation}
\end{equation}
Inserting Eq.~\eqref{eq_HankelTemperature} into Eq.~\eqref{eq_BCThermalEquation} provides the expressions of $T_1$ et $T_2$ 
\begin{equation}
T_{1,2} =  \frac{1}{\chi(1-\gamma^2 \chi^2)} {\left [ \frac{e^{-\left(\frac{1}{\gamma}\mp\chi\right) h} - e^{\left(\frac{1}{\gamma}\mp\chi\right) h}}{e^{2\chi h}-e^{-2\chi h}} \right ]}.
\label{eq_ConstantsT1T2}
\end{equation}
The temperature field is then considered as a source term in the elastic wave equation. The displacement field $\mathbf{u}$ is derived from scalar $\phi$ and vector ${\Psi}$ potentials using Helmoltz decomposition. As the problem is cylindrical symmetric, the potentials $\phi$, and ${\Psi}=(0, 0,\psi)$ can be used to write
\begin{equation}
\mathbf{u} = {\nabla}\phi + {\nabla} \times {\nabla} \times {\Psi}.
\label{eq_PotentialDecomposition}
\end{equation}
This comes from the fact that in cylindrical coordinates $(r,\theta,z)$, the rotational is $\nabla \times {\Psi}=(0,-\partial \psi/ \partial r,0)$. Then, the wave equation results in two uncoupled equations for scalar and vector potentials
\begin{equation}
{\left\lbrace
\begin{array}{rcl}
(\lambda+2\mu)\nabla^2 \phi -  \rho\frac{\partial^2\phi}{\partial t^2} & = & \alpha_T(3\lambda + 2\mu) T,
\\
\mu \nabla^2\psi - \rho\frac{\partial^2\psi}{\partial t^2} & = & 0,
\end{array}
\right.}
\label{eq_PotentielWaveEquations2}
\end{equation}
where $\lambda$ and $\mu$ are the Lam\'e constants, and $\alpha_T$ is the coefficient of linear thermal expansion. Introducing $\eta = \alpha_T(3\lambda + 2\mu)/(\lambda+2\mu)$ and the bulk elastic velocities $c_L$ and $c_T$, the following system is obtained by Fourier-Hankel transform
\begin{equation}
{\left\lbrace
\begin{array}{lr}
\frac{\partial^2 \overline{\phi}^{H_0} }{\partial z^2} - p^2 \overline{\phi}^{H_0} = \eta \overline{T}^{H_0}
&{\rm  \quad with \quad } p^2 = k^2 - \frac{\omega^2}{c_L^2},\\ \\
\frac{\partial^2 \overline{\psi}^{H_0} }{\partial z^2} - q^2 \overline{\psi}^{H_0}=0
&{\rm \quad with \quad } q^2 = k^2 - \frac{\omega^2}{c_T^2}.
\end{array}
\right.}
\label{eq_BesselFourierDiffSyst}
\end{equation}
The potentials $\overline{\phi}^{H_0}$ and $\overline{\psi}^{H_0}$ are the sum of a particular solution and a solution of the homogeneous equation. They can be written in the form
\begin{widetext}
\begin{equation}
{\left\lbrace
\begin{array}{lll}
\overline{\phi}_{H_0} =& \left[ \mathcal{A}e^{p z} + \mathcal{B}e^{- p z} + \left(\frac{1}{\chi^2-p^2}\right) (T_1e^{\chi z} + T_2e^{- \chi z})-\left(\frac{\gamma^2}{1-\gamma^2 p^2}\right) \left(\frac{\gamma}{1-\gamma^2 \chi^2}\right)e^{-\frac{z}{\gamma}}\right] \frac{\eta}{K} E^{H_0}(k)e^{-\frac{h}{\gamma}},  \\
\overline{\psi}_{H_0} =& \left[ \mathcal{C}e^{q z} + \mathcal{D}e^{- q z}\right] \frac{\eta}{K} E^{H_0}(k)e^{-\frac{h}{\gamma}},\\
\end{array}
\right.} 
\label{eq_PotentialSolutions}
\end{equation}
\end{widetext}
where the constants $\mathcal{A}$, $\mathcal{B}$, $\mathcal{C}$ and $\mathcal{D}$ are determined from boundary conditions as explained in the Appendix~\ref{secA}.
For each mode, the displacement components of $\overline{u}(k,z,\omega)$ are simply expressed as a linear combination of the potentials [Eq.~\eqref{eq_HankelFourierPotentialToDisplacementEquations}]. Consequently, the normal displacement at a given $(\omega_0,k_0)$ is proportional to the Hankel transform of the source spatial distribution $E^{H_0}(k_0)$. In particular, a ZGV mode can be enhanced by optimization of the source Hankel transform at spatial frequency $k_0$. Furthermore, as explained in the Appendix [Eq.~\eqref{eq_displacement}], the mechanical displacement $u(r,z,\omega)$ can be calculated numerically by inverse Hankel transform.

\section{Derivation and simulation of the optimal beam parameters}\label{sec2}

We now consider different source geometries and derive the optimal parameters for the generation of a Lamb mode. As previously discussed and demonstrated in the Appendix, the displacement components can be written as 
\begin{equation}
{\left\lbrace
\begin{array}{lcl}
\displaystyle \overline{u}_r^{H_1} = U_{r}  E^{H_0}(k),  \\
\displaystyle \overline{u}_z^{H_0} = U_{z}  E^{H_0}(k), 
\end{array}
\right. }
\label{eq_DisplacementDependanceWithPowerDensity}
\end{equation}
where the functions $U_r$ and $U_z$ are independent of the source geometry. Considering a spatial distribution depending on a parameter $a$, if the amplitude of the mode at fixed $(\omega,k)$ undergoes maximum, then 
\begin{equation}
{ \left.  \frac{\partial E^{H_0}}{\partial a} \right|}_{k} = 0.
\label{eq_DistributionOptimizationCondition}
\end{equation}
Additionally, the maximal surface energy $I$ is supposed to remain below the ablation threshold $I_a$.

These two conditions are now applied to different beam geometries.

\subsection{Optimization of a Gaussian beam}

\begin{figure}[!ht]
\centering
\includegraphics[width=\columnwidth]{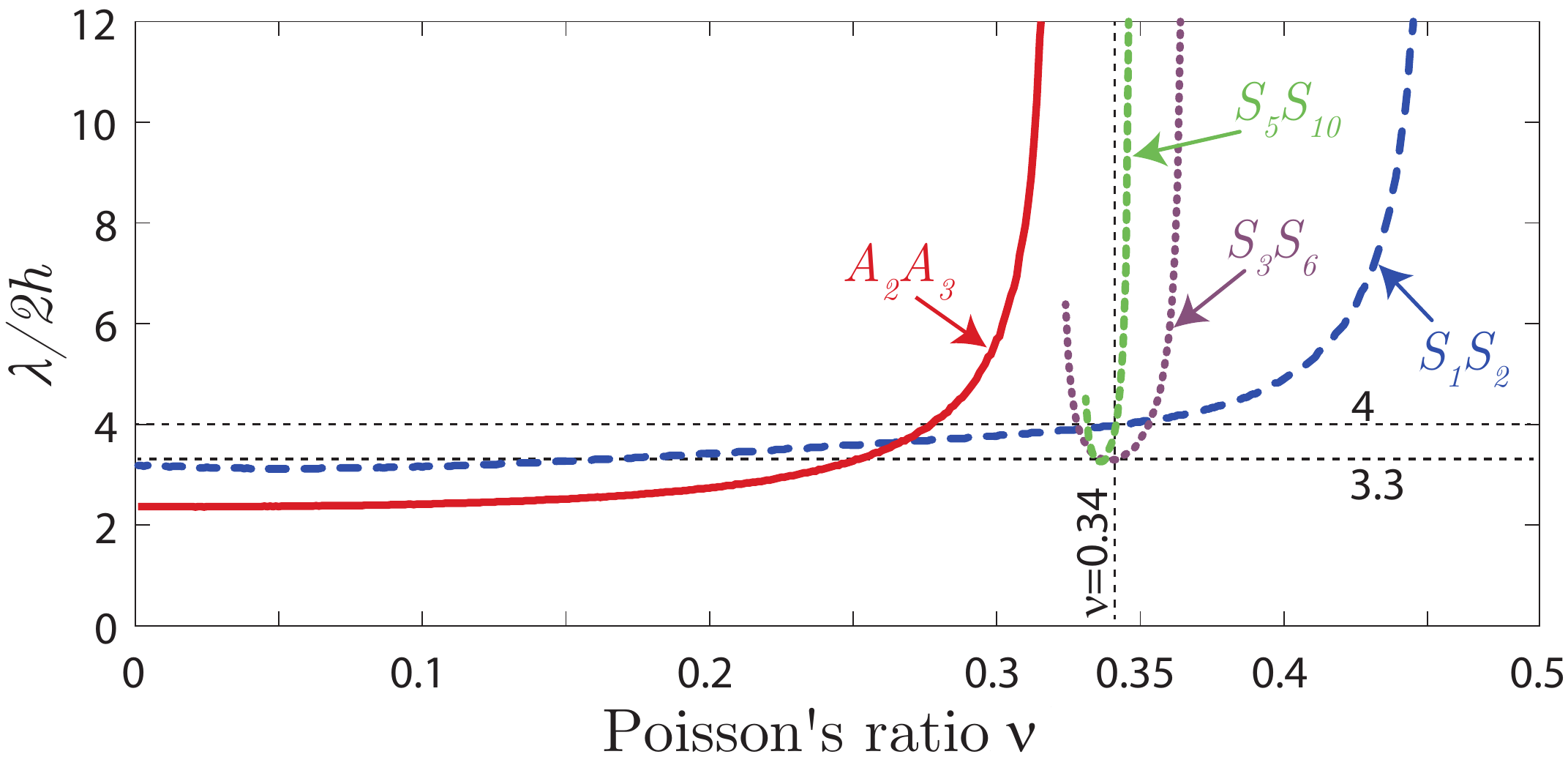}
\caption{ZGV modes wavelength versus the Poisson's ratio.}
\label{fig:LambdaVersusPoisson}
\end{figure}

The conditions equations~\eqref{eq_DistributionOptimizationCondition} and $I<I_a$ are first applied for a Gaussian source of radius $\mathrm{w}$. The absorbed energy distribution is written as $E(r) = I\exp(-r^2/\mathrm{w}^2)$. The total absorbed energy is $E_{tot} = \pi \mathrm{w}^2 I$ and the resulting Hankel transform is given by
\begin{equation}
E^{H_0}(k) = \frac{I\mathrm{w}^2}{2}e^{-\left(\frac{\mathrm{w}^2 k^2}{4}\right)}. \notag
\label{eq_NormalizedGaussianProfile}
\end{equation}
The amplitude of the mode at $(\omega ,k)$ reaches a maximum for an optimal beam radius equal to
\begin{equation}
\mathrm{w}_{opt} = 2/k =\lambda/\pi.
\label{eq_Gauss_Wopt}
\end{equation}
This very simple formula shows that the optimal waist only depends on the mode wavelength, which is proportional to the plate thickness and function of the elastic parameters. The wavelength of ZGV modes have been calculated as a function of the Poisson's ratio and are displayed in Fig.~\ref{fig:LambdaVersusPoisson}. For the $S_1S_2$-ZGV mode, the wavelength varies from about three times the plate thickness for hard materials $(\nu\approx 0)$ to four times the plates thickness for usual metals $(\nu\approx 0.3)$ and increases to infinity for the value $\nu=0.451$ where the ZGV point reaches the shear thickness resonance.

\begin{figure}[!ht]
\centering
\includegraphics[width=\columnwidth]{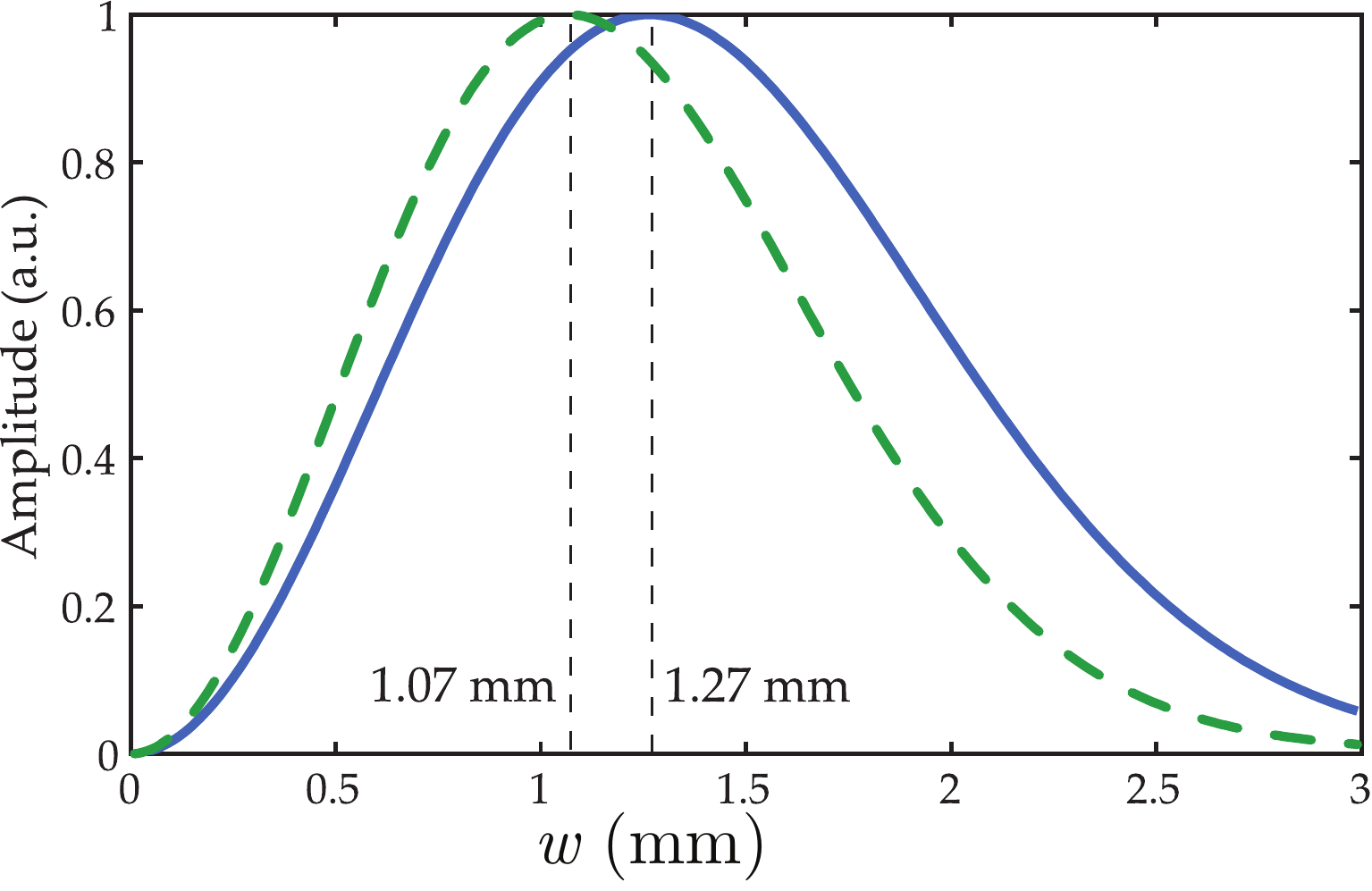}
\caption{Amplitude of the normal surface displacement as a function of the radius at $1/e$ of a Gaussian beam for $1$-\mm thick Duralumin (solid line) and fused silica (dashed line) plates. Theoretical optimal radii $\mathrm{w}_{dural} = 1.27$~\mm and $\mathrm{w}_{Silica} = 1.06$~\mm are in good agreement with the curves.}
\label{fig:GaussBeam_Simu_W_DuralFusedSilica}
\end{figure}

To illustrate the result given by Eq.~\ref{eq_Gauss_Wopt}, we consider the case of $1$-\mm thick Duralumin and fused silica plates. For Duralumin of bulk velocities $c_L = 6398$~m/s and $c_T = 3122$~m/s, the $S_1S_2$-ZGV mode wavelength is $\lambda_{S_1S_2} = 3.99$~\mm, while for fused silica ($c_L = 5961$~m/s, $c_T = 3727$~m/s) it is $3.33$~\mm. The theoretical optimal waists are then $\mathrm{w}_{dural} = 1.27$~\mm and $\mathrm{w}_{silica} = 1.06$~\mm. For both plates, the normal surface displacements at the ZGV frequency for $r=0$ are calculated as a function of the laser beam radius $\mathrm{w}$ [Fig.~\ref{fig:GaussBeam_Simu_W_DuralFusedSilica}]. The maxima of these curves are in good agreement with theoretical calculations.

These results are coherent with the approximation ($\mathrm{w}\approx 0.3\lambda_{S_1S_2}$) given for an aluminium plate by Balogun~\etal~\cite{BalogunMurrayPrada2007_JAP} Other results were obtained by Grünsteidl~\etal in a $250$-\um thick tungsten plate~\cite{Grunsteidl15} of Poisson's ratio $\nu = 0.284$ and shear velocity $c_s = 2668$~m/s. With these parameters, the wavelength is $\lambda_{S_1S_2} = 3.7 \times 2h$~\mm, so that the optimal beam radius to plate thickness ratio is
\begin{equation}
\frac{\mathrm{w}}{2h}=\frac{\lambda}{2\pi h}=1.17.
\label{Eq:blabla}
\end{equation}
To compare with their simulation $D/h \approx 1.5$ and measurement $D/h \approx 1.4$, one has to figure out, despite several inconsistencies in the paper, that $D$ was defined as the beam radius at $1/e^2$ : $D = \mathrm{w}\sqrt{2}$ and that $h$ is the plate thickness. With our notations, these ratio correspond to $\mathrm{w}/(2h)=1.06$ and $0.99$ respectively, which is in reasonable agreement with equation~\eqref{Eq:blabla}.

\subsection{ Optimization of an annular beam}

\begin{figure}[!ht]
\centering
\includegraphics[width=0.9\columnwidth]{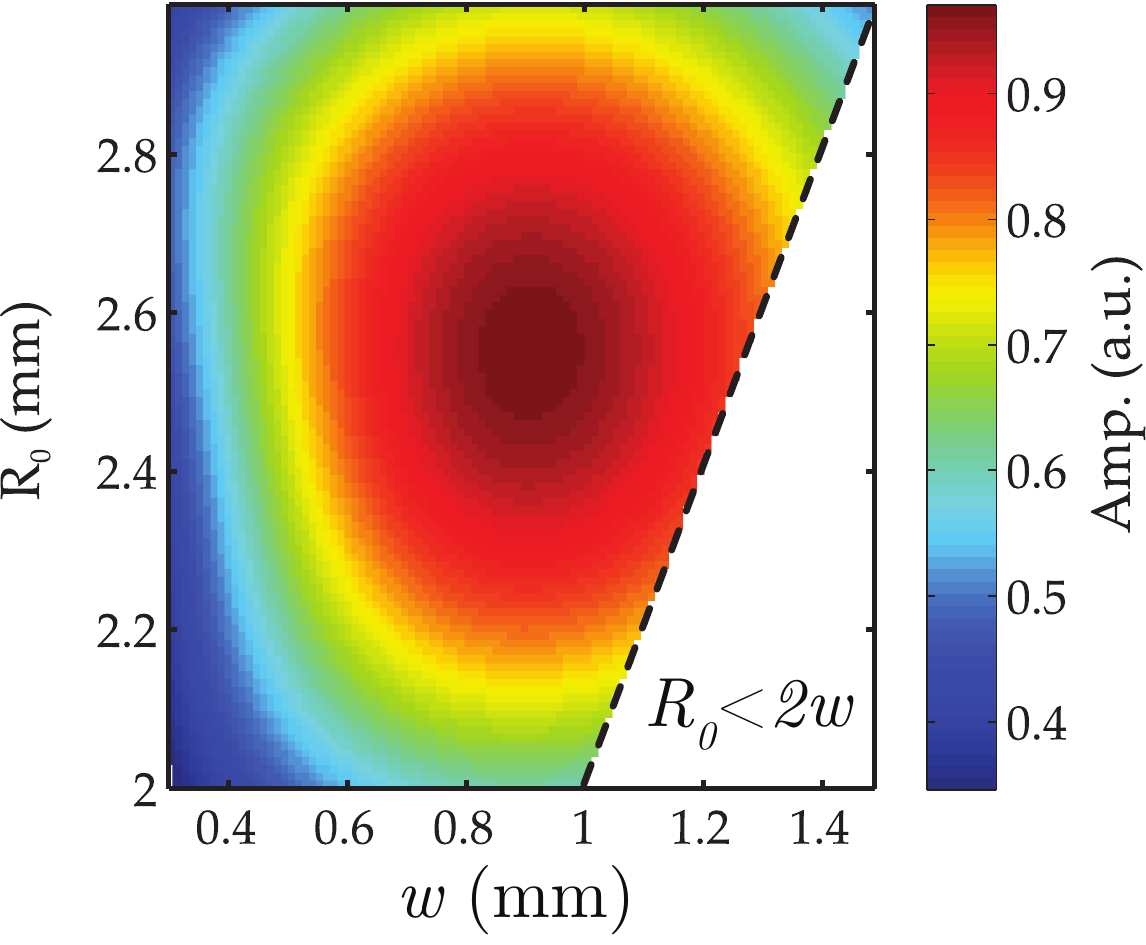}
\caption{Amplitude of the normal displacement as a function of radius $R_0$ and width $\mathrm{w}$ of a Gaussian ring in a 1-\mm thick Duralumin plate at constant peak energy. The maximum is reached for $\mathrm{w} = 0.88$~\mm and $R_0 = 2.57$~\mm which is in good agreement with the predicted values.}
\label{fig:GaussRing_Simu_R0Wmap}
\end{figure}

We now consider a ring with a Gaussian profile. Such spatial distribution depends on the radius $R$ and the half-width $\mathrm{w}$ as follows:
\begin{equation}
E(r) = Ie^{-{\left( \frac{r-R}{\mathrm{w}} \right)^2}}. \notag
\label{eq_TheoricalRingGaussianProfile}
\end{equation}
In order to derive analytical expressions, $E(r)$ is approximated by the convolution of a Gaussian function and a circular Dirac by using the normalized spatial distribution  
\begin{equation}
f(r) = \frac{1}{\pi \mathrm{w}^2} e^{-{\left( \frac{r}{\mathrm{w}} \right)}^2 } \ast  \frac{1}{2\pi R_0} \delta{\left( r-R_0 \right)}. \notag
\label{eq_ProfilDistributionApproxConv}
\end{equation}
A good approximation of $E(r)$ is obtained when the parameter $R_0$ is given by
\begin{equation}
R_0=R{\left(1 + \frac{\mathrm{w}^2}{4R^2} \right)}, \notag
\label{eq_ProfilDistributionLinkRR0}
\end{equation}
and when the radius is larger than the annular width, \ie, for $ R > 2\mathrm{w}$. It can be shown that the total absorbed energy is approximatively given by 
\begin{equation}
E_{tot}  \approx 2\pi^{3/2}\mathrm{w}R_0I.
\label{eq_EtotRing}
\end{equation}
The resulting Hankel transform of $f(r)$ is 
\begin{equation}
f^{H_0}(k) = \frac{1}{2\pi}e^{-\left(\frac{\mathrm{w}^2k^2}{4}\right)}J_0(k R_0).
\label{eq_HankelTransformRing}
\end{equation}
This function is separable so that the optimization can be performed independently on $\mathrm{w}$ and $R_0$,
\begin{equation}
\begin{matrix}
{\left. \frac{\partial}{\partial \mathrm{w}} {\Bigl(E_{tot} f^{H_0}(k) \Bigr)} \right|}_{k,R_0} = 0   &\Rightarrow&  \mathrm{w} = \lambda/(\pi\sqrt{2}), & \\ 
{\left. \frac{\partial}{\partial R_0} {\Bigl(E_{tot} f^{H_0}(k) \Bigr)} \right|}_{k,\mathrm{w}} = 0  &\Rightarrow&  \frac{\partial}{\partial R_0} {\Bigl(R_0 J_0(k R_0) \Bigr)}  =  0.
\end{matrix}
\label{eq_OptimisationRingR0_Etape1}
\end{equation}
As $R_0>2\mathrm{w}$, the asymptotic expansion of $J_0(k R_0)$ valid for $k R_0 = \sqrt{2}R_0/\mathrm{w}> 1/4$ can be used, 
\begin{equation}
R_0^{(n)} = \lambda\frac{(4n+1)}{16}{\left( 1+\sqrt{1+\frac{4^3}{(4n+1)^2\pi^2}} \right)},
\label{eq_OptimisationRingR0}
\end{equation}
where $n \geq 1$ is the order of the ring. Finally, taking into account Eq.~\eqref{eq_OptimisationRingR0_Etape1}, the previous assumption $R_0^{(n)}>2\mathrm{w}$ is always fulfilled for all orders $n$. In a 1-\mm thick Duralumin plate of Poisson's ratio $0.344$, the wavelength is $\lambda_{S_1S_2} = 3.99$~\mm. For the first ring, $n=1$, the optimized parameters are $\mathrm{w} = 0.90$~\mm, and $R_0 = 2.57$~\mm. The result of simulations, shown (Fig.~\ref{fig:GaussRing_Simu_R0Wmap}), provide optimal parameters that are in good agreement with the theoretical parameters.

The amplitude gain $G$ is defined as the ratio between the normal displacement $\overline{u}_z^{H_0}$ obtained with the \textrm{$n^{th}$} annular source [Eq.~\eqref{eq_DisplacementDependanceWithPowerDensity}] and that obtained with the optimized Gaussian beam
\begin{equation}
G =\left| {E^{H_0}_{ring}}/{E^{H_0}_{gaussian}}\right| \approx  (e/2)\sqrt{4n+1}.
\end{equation}
It turns out that the amplification factor expected for the first-order optimized ring ($n = 1$) is about $3$. The use of optimized ring of greater order will raise the amplification factor proportionally to the square root of the ring order. This is reasonable as the Bessel function $J_0(k r)$ decreases as $\sqrt{r}$ while the total of deposited energy increases proportionally to the ring radius as shown in Eq.~\eqref{eq_EtotRing}.

\subsection{Optimization of a top-hat source and a rectangular annular source}

The profile distribution for a top-hat source and the resulting Hankel transform are
\begin{equation}
E(r) = \Pi\left(\frac{r}{\mathrm{w}}\right)I
\quad\Rightarrow\quad 
E^{H_0}(k) = \frac{I\mathrm{w}}{k} J_1(\mathrm{w}k),
\label{eq_NormalizedRectRingProfile}
\end{equation}
where $\Pi(r)$ is the top-hat function defined as
\begin{equation}
\Pi(r) = 
\left\{
\begin{matrix}
 1, & {\rm if \quad} r \leqslant 1,\\ 
 0, & {\rm elsewhere}.
\end{matrix}\right.
\label{eq_diskFunction}
\end{equation}
The optimal $\mathrm{w}$ is such that 
\begin{equation}
{\left. \frac{{\rm d}E^{H_0}}{{\rm d} \mathrm{w}} \right|}_{k}  
=\frac{\mathrm{w}}{k} J_0(\mathrm{w}k) = 0.
\end{equation}
It appears that for a rectangular profile, the amplitude maxima occur for several radii $\mathrm{w}$,
\begin{equation}
\mathrm{w}^{(n)} \approx \frac{\lambda}{2} {\left( n+\frac{3}{4}  \right)},
\end{equation}
and the corresponding amplification factors compared to the optimized Gaussian spot are approximated as 
\begin{equation}
G = {\left| {E^{H_0}_{\Pi}}/E^{H_0}_{gaussian} \right|} \approx  (e/2)\sqrt{n +(3/4)}.
\end{equation}
For $n=0$, this amplification factor is about $1.7$. It is remarkable that, unlike the Gaussian beam, the amplitude maxima increase with the spot radius.\\

The intensity profile for a rectangular ring is written
\begin{equation}
E(r) = I\left[\Pi\left(\frac{r}{R_0+\mathrm{w}}\right)-\Pi\left(\frac{r}{R_0-\mathrm{w}}\right)\right]. \notag
\label{eq_NormalizedRectRingProfile}
\end{equation}
The associated total energy is $E_{tot} = 4\pi \mathrm{w} R_0 I$ and the resulting Hankel transform of $E(r)$ is equal to
\begin{align}
E^{H_0}(k) &= \frac{I}{k^2}\Bigl( (R_0+\mathrm{w})kJ_1[(R_0+\mathrm{w})k] \notag\\ 
           &- (R_0-\mathrm{w})kJ_1[(R_0-\mathrm{w})k] \Bigr).  
\label{eq_HankelTransformNormalizedRectRingProfile}
\end{align}
The derivative with respect to $\mathrm{w}$ vanishes for 
\begin{equation}
{\left\lbrace
\begin{array}{lllll}
R_0-\mathrm{w} &=& \frac{\lambda}{2} {\left( n + \frac{3}{4} \right)}, && n \in \mathbb{N}^*,   \notag\\
R_0+\mathrm{w} &=& \frac{\lambda}{2} {\left(m + \frac{3}{4} \right)},  && m>n \in \mathbb{N}^*. \notag 
\end{array}
\right.}
\end{equation}
Consecutive solutions with $m=n+1$ lead to 
\begin{equation}
{\left\lbrace
\begin{array}{lclcr}
R_0^{(n)}  &=& \frac{\lambda}{8}\left( 4n +1 \right), & & n \in \mathbb{N}^*, \\
\mathrm{w} &=&  \lambda/4.
\end{array}
\right.}
\end{equation}
The amplitude gain obtained with the $n^{th}$ rectangular ring compared to the optimal Gaussian spot is provided by
\begin{align}
G &=  \left| E^{H_0}_{rect \ ring}/{E^{H_0}_{gaussian}}\right|\nonumber\\
  & \approx  e\sqrt{n/2} {\left( \sqrt{1+3/(4n)} + \sqrt{1-1/(4n)} \right)}.  
\label{eq_AmplificationFactor_RectRing}
\end{align}
For the first-order optimized ring ($n=1$), the amplification factor is about $4.2$ which is significantly higher than the Gaussian ring. The optimal parameters, associated with total deposited energies and the amplitudes for the different source shapes ($n=1$ for the rings) are gathered in Table~\ref{table:Resume}. It appears that the amplitude gain $G$ for top-hat beam is higher than the gain in the total energy given by $\widetilde{E}_{tot}$. Similar observation arises from the comparison of Gaussian and rectangular rings. This can be ascribed to the strong temperature gradients that induce high in-plane constrains at the beam edge and reinforces the idea that it would be advantageous to use rectangular energy profiles.\\

\begin{table}[!ht]
\begin{minipage}[c]{\columnwidth}
\begin{ruledtabular}
\centering
\caption{ Theoretical optimized geometric parameters $\mathrm{w}$ and $R_0^{(1)}$, total absorbed energy normalized to the Gaussian case ($\widetilde{E}_{tot}$) and gain $G$ calculated for the $S_1S_2$-ZGV mode in a 1-\mm thick Duralumin plate and different source shapes.}
\begin{tabular}{lccccc}
\textbf{Source shape} & $\mathrm{w}$ (\mm) & $R_0^{(1)}$ (\mm) &  $\widetilde{E}_{tot}$ & $G$ \\
\hline
Gaussian beam 		& $ 1.27 $ & $ NA  $ &  $ 1    $ & $ 1    $ \\
Top-hat 			& $ 1.50 $ & $ NA  $ &  $ 1.39 $ & $ 1.66 $ \\
Gaussian ring 		& $ 0.90 $ & $ 2.57 $ &  $ 5.08 $ & $ 3.27 $ \\
Rectangular ring 	& $ 1.00 $ & $ 2.49 $ &  $ 6.17 $ & $ 4.21 $ \\
\end{tabular}
\label{table:Resume}
\end{ruledtabular}
\end{minipage}
\end{table}

\begin{figure}[!ht]
\centering
\includegraphics[width=\columnwidth]{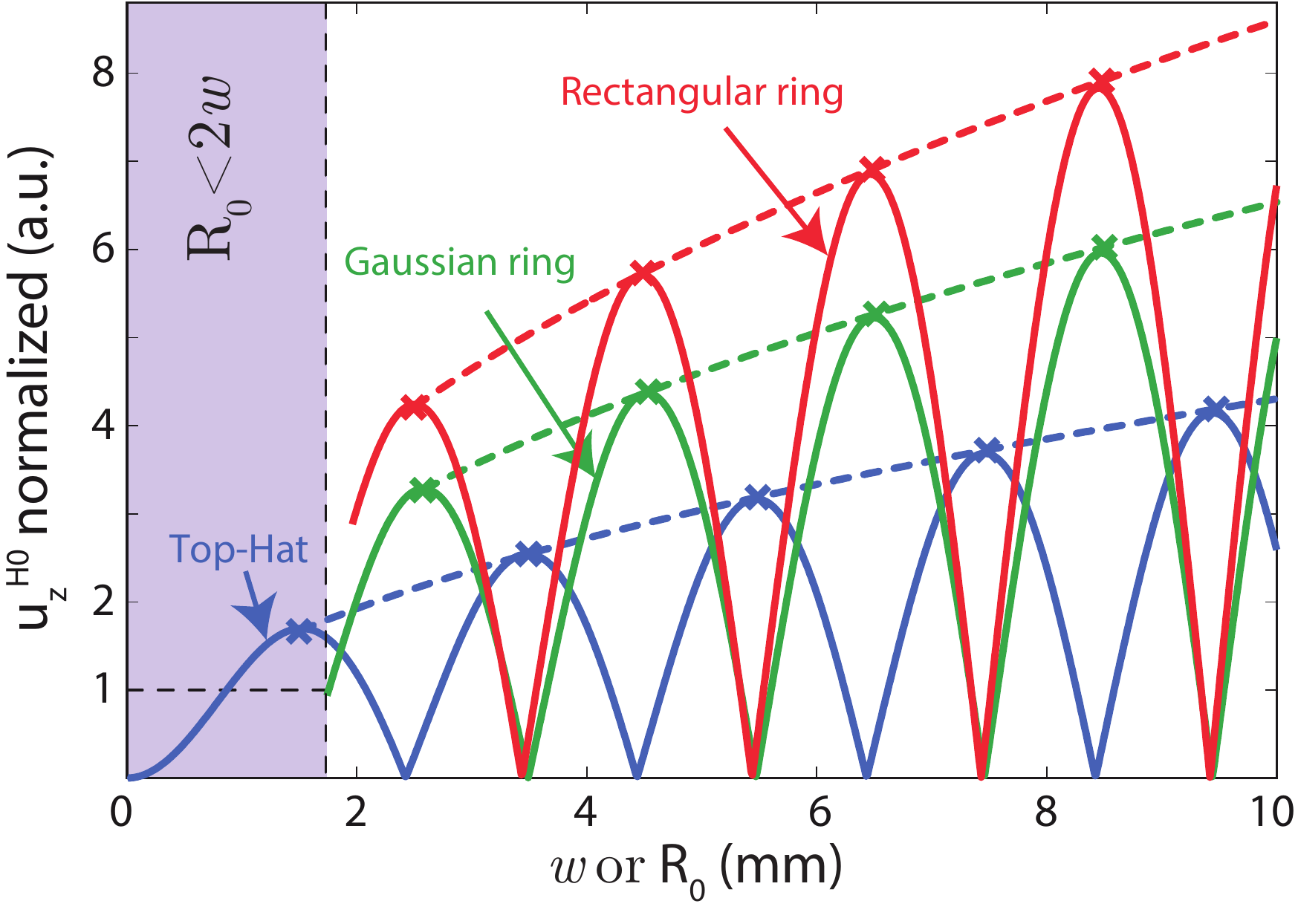}
\caption{Displacement amplitude at $S_1S_2$-ZGV frequency in a $1$-\mm thick Duralumin plate for top-hat profile as a function of $\mathrm{w}$, and for Gaussian and rectangular rings as a function of the radius $R_0$ for a fixed optimal width $\mathrm{w}$. Curves are normalized to the displacement obtained with the optimal Gaussian beam.} 
\label{fig:SimuVsTheory_Rn_AllExcitation}
\end{figure}

\begin{figure}[!ht]
\centering
\includegraphics[width=\columnwidth]{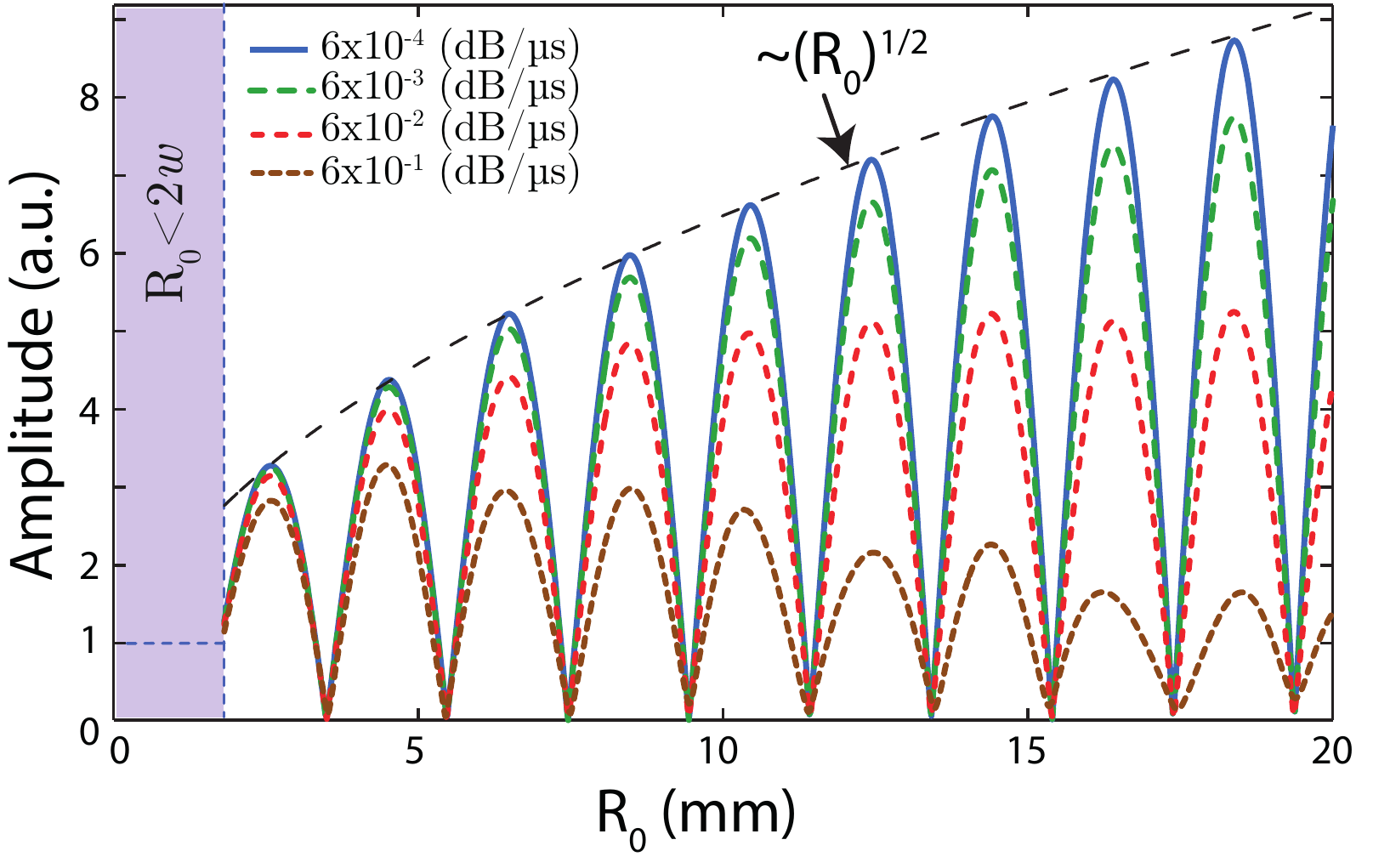}
\caption{Displacement amplitude for Gaussian ring sources of width $\mathrm{w} = 0.88$~\mm as a function of $R_0$ calculated for different damping parameters.}
\label{fig:GaussRing_Simu_R0n_AttenuationEffect}
\end{figure}

From Eq.~\eqref{eq_AmplificationFactor_RectRing}, it appears that the factor $ G $ is proportional to the square-root of the ring order. The amplitude calculated with the semi-analytical simulation as a function of the ring radius $R_0$ is plotted in Fig.~\ref{fig:SimuVsTheory_Rn_AllExcitation}. These theoretical results suggest that high order annular beam should be used to obtain maximal amplitude. However, they are correct only for low attenuating material. In general, the larger is the ring, the more important is the effect of attenuation. This can been observed on the displacement amplitude, simulated for Gaussian rings of optimal width and different damping parameters (Fig.~\ref{fig:GaussRing_Simu_R0n_AttenuationEffect}). 
A damping factor of $6\times 10^{-4}$ \dbus corresponds to a lossless material, the second one to weakly attenuating materials like Duralumin, the third one to steel or copper and the value $6\times 10^{-1}$ \dbus to highly attenuating materials like composite plates. For a damping factor of $6 \times 10^{-2}$ \dbus, attenuation cannot be neglected, the gain between the Gaussian beam and the first ring is around three but it increases slowly for higher orders rings. Whatever the attenuation, the first annular ring provides a significant gain.

\section{Experimental Results}\label{sec3}

\begin{figure*}[!ht]
\centering
\subfigure{\includegraphics[width=0.8\textwidth]{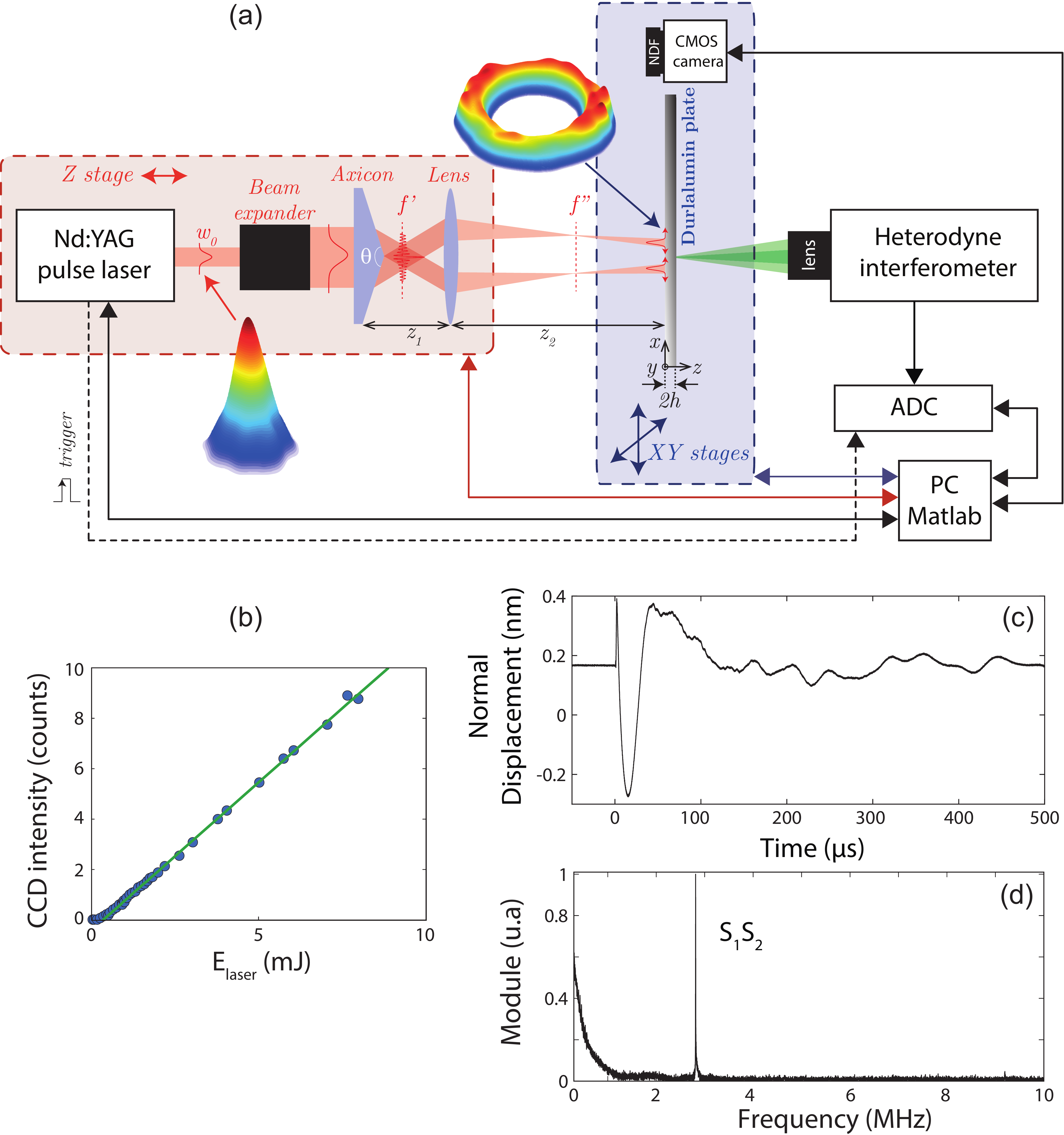}\label{fig:SchemaSetUpMeasurea}}
\subfigure{\label{fig:SchemaSetUpMeasureb}} 
\subfigure{\label{fig:SchemaSetUpMeasurec}}
\subfigure{\label{fig:SchemaSetUpMeasured}} 
\caption{(a) Experimental set-up. (b) Total energy measured on the camera as a function of the laser beam energy. (c) Typical measured normal displacement, and (d) associated spectrum.}
\label{fig:SchemaSetUpMeasure}
\end{figure*}

The experimental set-up is shown in Fig.~\ref{fig:SchemaSetUpMeasurea}. The excitation is achieved with a Nd:YAG laser at $1064$ \nm (Centurion Quantel, pulse duration $8$ \ns and fire rate $100$ \Hz) and an optical system. Normal surface displacements are measured with a heterodyne interferometer (BMI probe, SH-140, calibration factor $120$~mV/nm).\cite{RoyerDieulesaint1986_APL} The signals detected by the optical probe were fed into a digital sampling board (PicoScope, PicoTech, 6404D) with $5$~ns resolution and then stored in a computer for further analysis. Two kinds of optical systems are used to produce either a Gaussian or a ring-shaped source. The first one consists in a beam expander with a $100$-mm convergent lens. The beam radius is controlled by varying the distance from the lens to the sample. The second system is composed of a beam expander, an axicon (\ie, a conical lens) of apex angle $\theta=160^{\circ}$, and a $35$-mm convergent lens. Radius and width of the annular source are controlled by varying the distances $z_1$, $z_2$ between the axicon, the lens, and the sample.\\

As shown in Fig.~\ref{fig:SchemaSetUpMeasurea}, the laser source is characterized with a camera (uEye, IDS, UI-3370CP, with $2048 \times 2048$ pixels of size $5.5$~\umm). A neutral density filter is placed just before the camera in order to avoid saturation on the CMOS sensor. Integration time is about $47$~\ms and pulse rate is $100$~\Hz. Sixty-four images are consecutively recorded and averaged. Under these conditions, the number of photo-electrons (counts) measured is proportional to the pulse energy of the laser source [Fig.~\ref{fig:SchemaSetUpMeasureb}]. The normal displacement generated with an annular beam and the associated spectrum, displayed in Figs.~\ref{fig:SchemaSetUpMeasurec} and \ref{fig:SchemaSetUpMeasured}, show that the $S_1S_2$ resonance is favourably excited.\\

\begin{figure}[!ht]
\centering
\subfigure{\includegraphics[width=0.75\columnwidth]{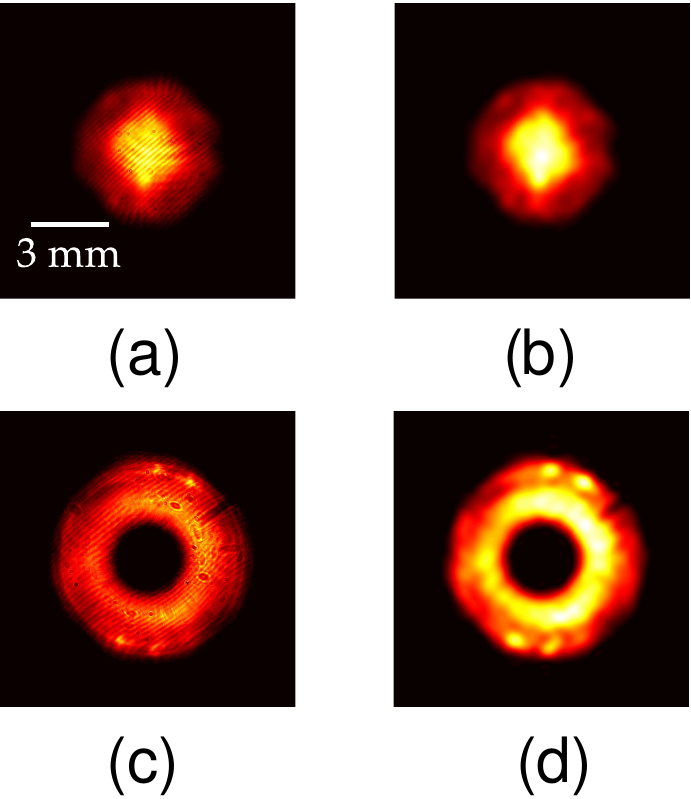}\label{fig:Image_GaussianSpot_GaussianRing_ImagesPreTreatmenta}}
\subfigure{\label{fig:Image_GaussianSpot_GaussianRing_ImagesPreTreatmentb}} 
\subfigure{\label{fig:Image_GaussianSpot_GaussianRing_ImagesPreTreatmentc}}
\subfigure{\label{fig:Image_GaussianSpot_GaussianRing_ImagesPreTreatmentd}} 
\caption{Snapshots of sources. (a) Raw Gaussian beam, and (b) after 2-D Fourier transform filtering. (c) Raw Gaussian ring, and (d) after 2-D Fourier transform filtering.}
\label{fig:Image_GaussianSpot_GaussianRing_ImagesPreTreatment}
\end{figure}

Examples of raw images are shown in Fig.~\ref{fig:Image_GaussianSpot_GaussianRing_ImagesPreTreatmenta} for a Gaussian beam and in Fig.~\ref{fig:Image_GaussianSpot_GaussianRing_ImagesPreTreatmentc} for an annular source. These images display linear fringes that are attributed to a glass plate in front of the CMOS sensor. In order to compare with theory, these interferences are suppressed by low pass filtering [Fig.~\ref{fig:Image_GaussianSpot_GaussianRing_ImagesPreTreatmentb} and \ref{fig:Image_GaussianSpot_GaussianRing_ImagesPreTreatmentd}]. The energy distribution is not exactly circular because of the imperfections of the laser source. Mean profiles are calculated by averaging 16 profiles, then they are fitted by functions
$e^{-r^2/\mathrm{w}^2}$ (for the Gaussian beam) and $e^{-(r-R)^2/\mathrm{w}^2} $ (for the annular beam). Once geometrical source parameters estimated, the maximum surface energy density $I$ is deduced from the normalized function $f(r)$ and measured total energy $E_{tot}$ by using Eq.\ref{eq_EnergyMax}.\\

\begin{figure}[!ht]
\centering
\includegraphics[width=\columnwidth]{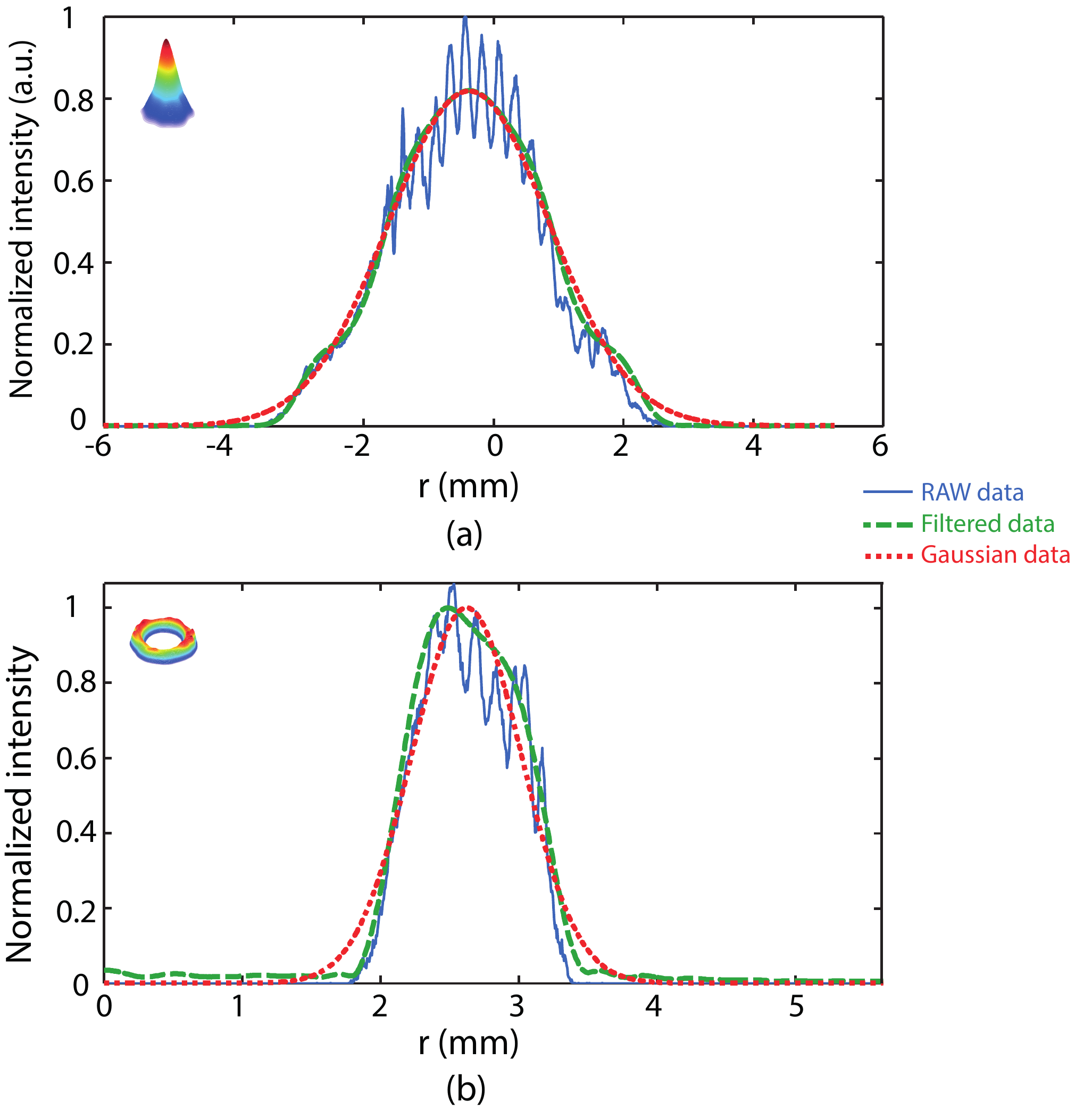}
\caption{Typical experimental profiles of the Gaussian (a) and annular (b) beams, and comparison with fitted profiles.}
\label{fig:ProfilsExpAndFit_GaussianSpot_GaussianRing}
\end{figure}

\begin{figure}[!ht]
\centering
\includegraphics[width=\columnwidth]{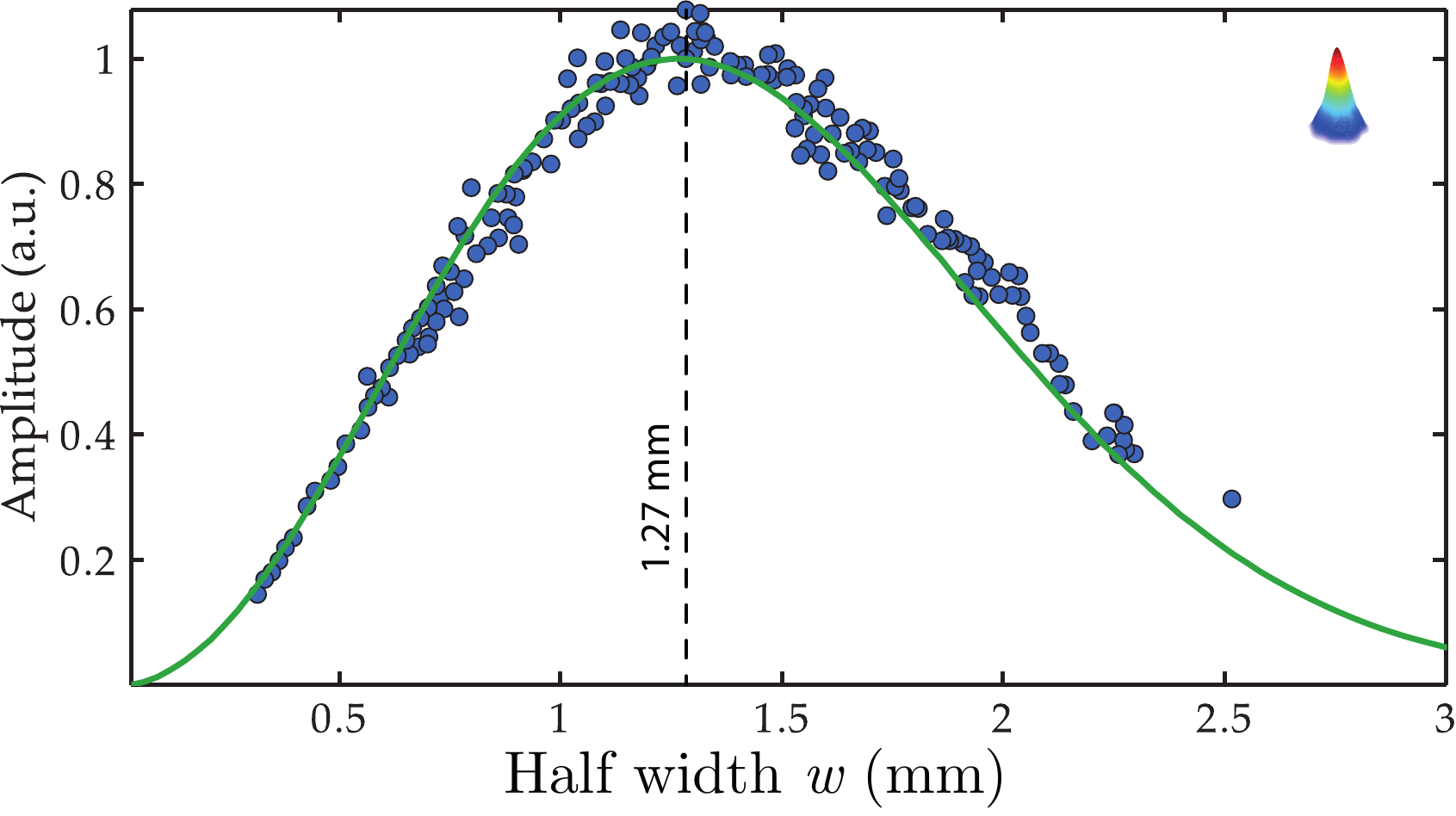}
\caption{Gaussian beam: Amplitude of $S_1S_2$-ZGV mode generated in the $1$-\mm thick Duralumin plate against the beam radius. Comparison between experiments (dots) and simulation (solid line).}
\label{fig:GaussianBeam_Amp_TheoVsExp}
\end{figure}

\textit{Gaussian beam} --- In order to validate the simulation and the experimental procedure, a first optimization is performed on a Gaussian laser source. All measurements are performed on a $1$-mm thick Duralumin plate. The beam radius is varied from $0.3$~mm to $2.5$~mm, and is measured systematically with the camera. The amplitude of the displacement at the $S_1S_2$-ZGV frequency $f=2.87$~MHz is normalized by the estimated maximum surface energy density. It is represented against the beam radius in Fig.~\ref{fig:GaussianBeam_Amp_TheoVsExp}. The agreement with the simulated displacement is noticeable. The relative root-mean-square error was found to be $6\% $ and the optimal radius coincides perfectly with the predicted value $\mathrm{w}_{opt} =\lambda/\pi= 1.27$~mm.\\

\begin{figure}[!ht]
\centering
\includegraphics[width=\columnwidth]{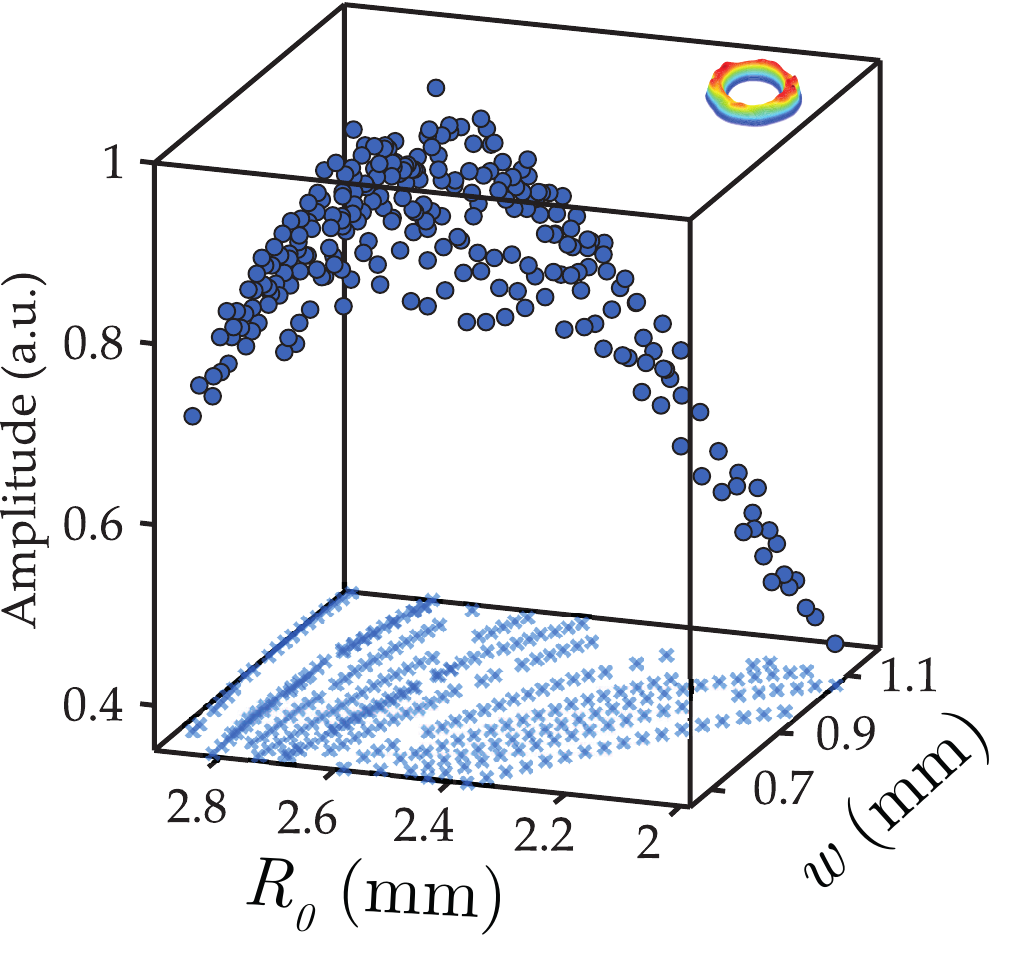}
\caption{Gaussian ring: Amplitude of the $S_1S_2$-ZGV mode generated in the $1$-\mm thick Duralumin plate against the ring width $\mathrm{w}$ and radius $R_0$. The crosses represent the set of $(\mathrm{w},R_0)$ geometric parameters.}
\label{fig:GaussRing_Exp_Dome}
\end{figure}

\begin{figure*}[!ht]
\centering
\subfigure{\includegraphics[width=0.8\textwidth]{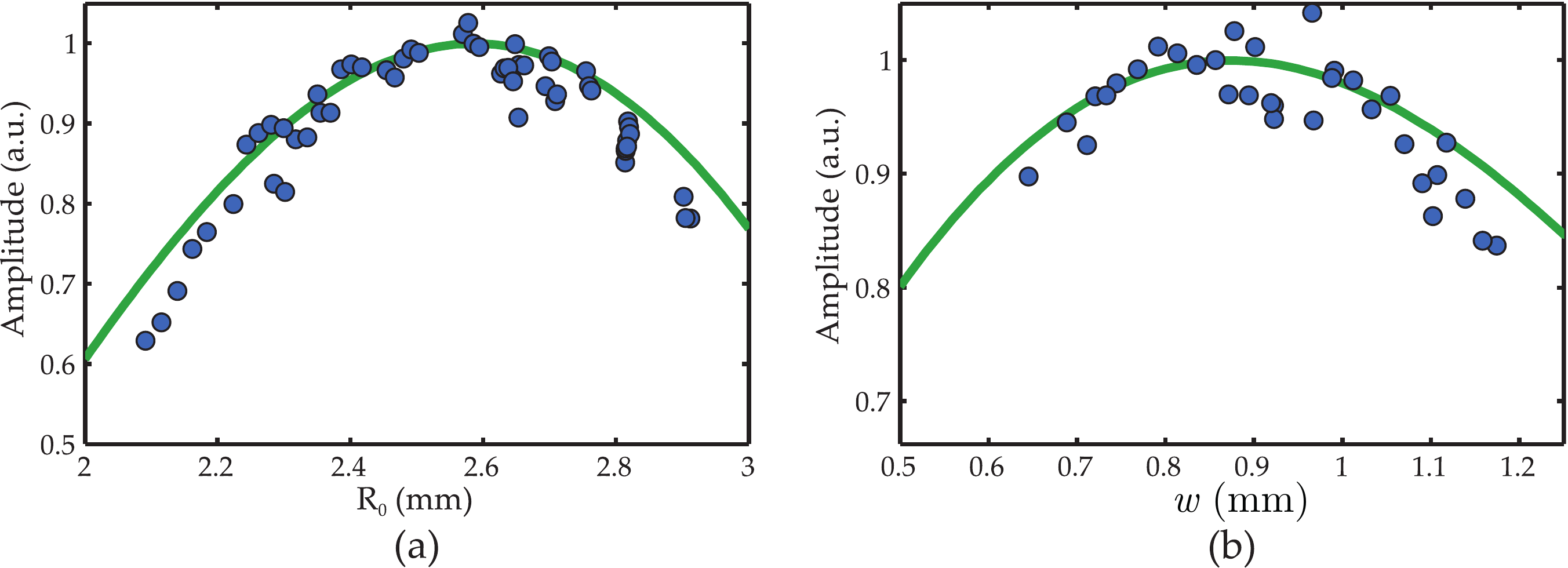}\label{fig:GaussRingExp_ComparisonToSimu_R0optAndWopta}}
\subfigure{\label{fig:GaussRingExp_ComparisonToSimu_R0optAndWoptb}} 
\caption{Gaussian ring: Experimental ZGV amplitudes (dots) and simulation (solid line): (a) Simulation is performed at fixed width $\mathrm{w}_{opt}$ while experimental dots are chosen so that $|\mathrm{w}-\mathrm{w}_{opt}|<0.05$~\mm. (b) Simulation is performed at fixed radius $R_{0_{opt}}$, while experimental dots are chosen so that $|R_0 - R_{0_{opt}}|<0.05$~\mm.}
\label{fig:GaussRingExp_ComparisonToSimu_R0optAndWopt}
\end{figure*}

\textit{Annular beam} --- In practice, the width and the radius of the annular source cannot be changed independently. Axicon to lens distance is held constant while the lens to sample distance is incremented by $1$-\mm steps. Simultaneously, the geometric parameters of the source are measured and the displacement at the $S_1S_2$-ZGV resonance frequency is recorded. This procedure was repeated for a dozen axicon-lens distances covering the range from $2.0$ to $2.9$~\mm for $R_0$ and from $0.65$ to $1.20$~\mm for $\mathrm{w}$. These ranges are centered around the predicted optimal parameters given in Table~\ref{table:Resume}. Displacement amplitudes normalized by the estimated maximum surface energy density are plotted in Fig.~\ref{fig:GaussRing_Exp_Dome}. Experimental values are compared with simulations in Fig.~\ref{fig:GaussRingExp_ComparisonToSimu_R0optAndWopt}. 
In Fig. \ref{fig:GaussRingExp_ComparisonToSimu_R0optAndWopta}, $R_0$ varies from $2.1$ to $2.9$~\mm for ring width limited to $\pm 0.05$~\mm around the optimal value $0.90$~\mm. Similarly in Fig. \ref{fig:GaussRingExp_ComparisonToSimu_R0optAndWoptb}, $\mathrm{w}$ varies from $0.65$ to $1.20$~\mm while the ring radius is limited to $\pm 0.05$~\mm around the optimal value $2.57$~\mm. A reasonable agreement between the whole experimental data set and the simulation is observed with a relative root-mean-square error equal to $5\%$.\\

\begin{figure}[!ht]
\centering
\includegraphics[width=\columnwidth]{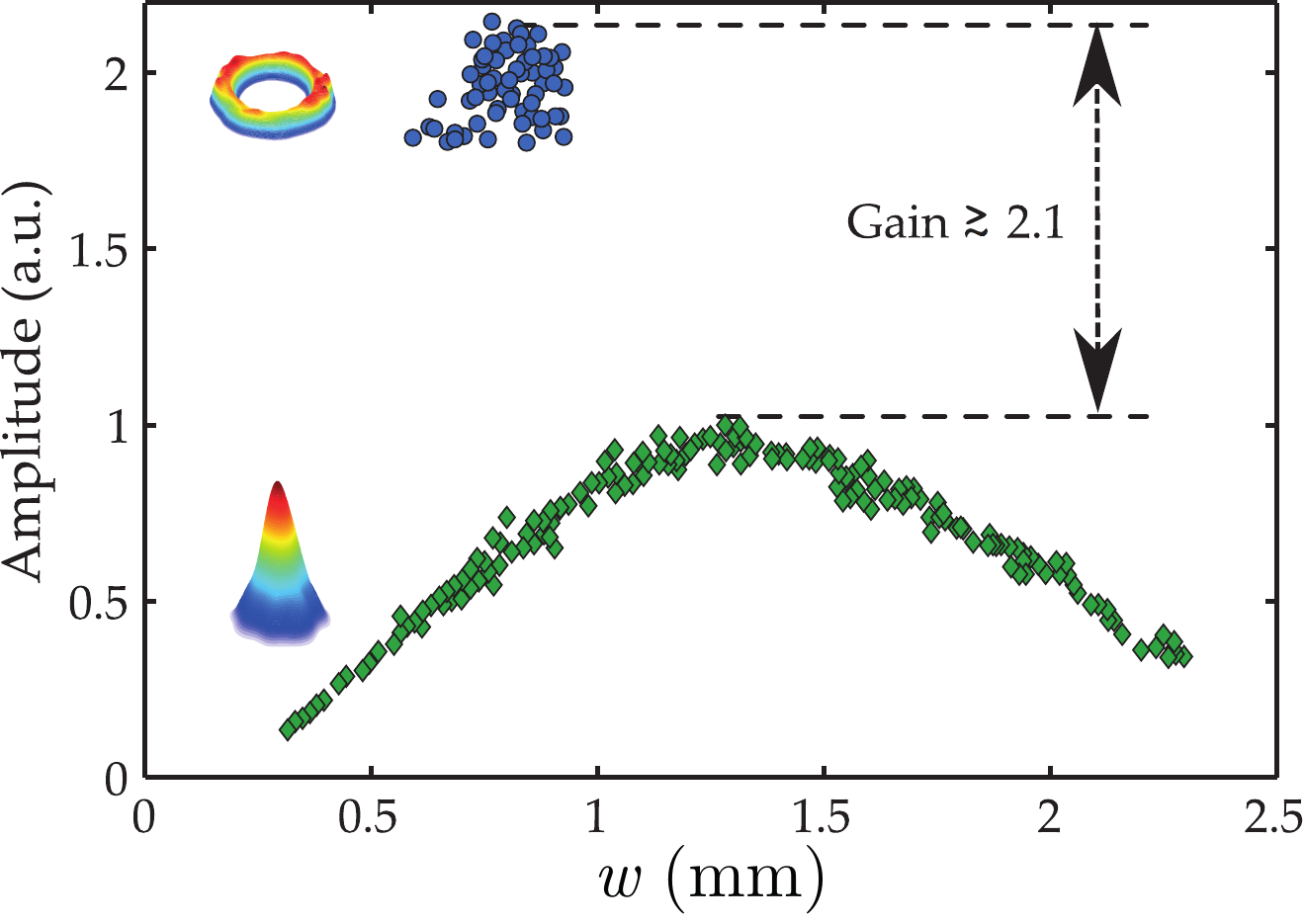}
\caption{Comparison of the displacement amplitude measured with Gaussian ring (blue dot) and Gaussian beam (green diamond).}
\label{fig:GaussRing_Exp_ComparaisonRingVsSpot}
\end{figure}

\textit{Gaussian beam versus Gaussian ring} --- Displacements measured experimentally are compared between the two sources in Fig.~\ref{fig:GaussRing_Exp_ComparaisonRingVsSpot}. The observed amplification factor ($2.1$) is significantly lower than the predicted one ($3.27$, Table~\ref{table:Resume}). This can be ascribed to various defaults: (i) the spatial distribution of the Gaussian laser is not axially symmetric, (ii) the energy profile of the annular source is not exactly Gaussian, (iii) the estimation of the geometric parameters of the annular source is difficult.\\

\section{Conclusion}
We have presented a theoretical study on Lamb modes generated by axi-symmetrical laser source. The objective was to improve the excitation of ZGV resonances in an isotropic plate by adjusting the geometric parameters of a Gaussian beam or an annular source. The first and simple result is that the optimal radius of a Gaussian beam is directly proportional to the ZGV wavelength and equal to $\lambda_{ZGV}/\pi$. Analytical formulas giving optimal parameters have also been established for Gaussian and rectangular rings. These theoretical results were confirmed by semi-analytical simulations. Annular Gaussian beams of controlled width and radii were achieved with an axicon-lens system. Amplitude of the $S_1S_2$ ZGV mode, measured for Gaussian spot and ring source in a Duralumin plates are in good agreement with theoretical predictions. These results obtained for ZGV modes, are also valid for the generation of any Lamb mode in a cylindrical geometry. Further works will consider the use of SLMs to shape optimal beams with a great flexibility. For non-destructive testing applications, tunable acoustic gradient (TAG) index lens could be a cheap and efficient alternative.\\

\section*{Acknowledgments}
This work was supported by LABEX WIFI (Laboratory of Excellence ANR-10-LABX-24) within the French Program ``Investments for the Future'' under reference ANR-10-IDEX-0001-02 PSL$^{\ast}$ and by the Chaire SAFRAN ESPCI.\\

\appendix
\section{Determination of the elastic potential and mechanical displacements}\label{secA}

Stress free surface conditions are applied in order to determine the constant $\mathcal{A}$, $\mathcal{B}$, $\mathcal{C}$, and $\mathcal{D}$ of the potentials in equation~\eqref{eq_PotentialSolutions}. Using $\rm{u} = {\nabla}\phi + {\nabla} \times {\nabla} \times {\psi}$ the displacement components are
\begin{equation}
{\left\lbrace
\begin{array}{lcl}
u_r & = & \frac{\partial \phi}{\partial r} + \frac{\partial^2\psi}{\partial r \partial z},\\
u_z & = & \frac{\partial \phi}{\partial z} + \frac{\partial^2\psi}{\partial z^2} - \frac{1}{c_T^2}\frac{\partial^2\psi}{\partial t^2},
\end{array}
\right.}
\label{eq_PotentialToDisplacementEquations}
\end{equation}
and the elastic stress components can be written as
\begin{equation}
{\left\lbrace
\begin{array}{lcl}
\sigma_{rz} & = & \mu \frac{\partial}{\partial r} {\left ( 2\frac{\partial \phi}{\partial z} + 2 \frac{\partial^2 \psi}{\partial z^2} - \frac{1}{c_T^2} \frac{\partial^2\psi}{\partial t^2} \right )}, \\
\sigma_{zz} & = & \lambda \nabla^2 \phi + 2\mu \frac{\partial}{\partial z} {\left ( \frac{\partial \phi}{\partial z} +  \frac{\partial^2 \psi}{\partial z^2} - \frac{1}{c_L^2} \frac{\partial^2\psi}{\partial t^2} \right )} -\eta(\lambda + 2\mu)T.
\end{array}
\right.} \notag
\label{eq_StressEquations}
\end{equation}
From Eq.~\eqref{eq_PotentialToDisplacementEquations}, the Hankel transforms of the displacement components can be expressed in terms of the potentials $\overline{\phi}^{H_0}$ and $\overline{\psi}^{H_0}$ as:
\begin{equation}
{\left\lbrace
\begin{array}{lcl}
\overline{u}_r^{H_1} = - k \overline{\phi}^{H_0} -k \frac{\partial \overline{\psi}^{H_0}}{\partial z}, \\
\overline{u}_z^{H_0}  = \frac{\partial \overline{\phi}^{H_0}}{\partial z} + k^2 \overline{\psi}^{H_0}, 
\end{array}
\right.}
\label{eq_HankelFourierPotentialToDisplacementEquations}
\end{equation}
where $^{``H_1"}$ denotes the Hankel transform of the first order. Then using the potential equations~\eqref{eq_BesselFourierDiffSyst} the stress components can be written as
\begin{equation}
{\left\lbrace
\begin{array}{lcl}
\overline{\sigma}_{rz}^{H_1} & = & - \mu k { \left ( \frac{2 \partial \overline{\phi}^{H_0}} {\partial z} + (k^2 + q^2) \overline{\psi}^{H_0} \right ) }, \\
\overline{\sigma}_{zz}^{H_0} & = & (-\lambda k^2 + (\lambda + 2\mu)p^2)\overline{\phi}^{H_0} + 2\mu k^2 \frac{\partial \overline{\psi}^{H_0}} {\partial z} .
\end{array}
\right.} \notag
\label{eq_HankelFourierStress}
\end{equation}
The elastic boundary conditions result from the absence of normal stress on both surfaces $z = \pm h$
\begin{equation}
{\left\lbrace
\begin{array}{lcl}
{\left . \sigma_{rz} \right |}_{z=\pm h} = 0\\
{\left . \sigma_{zz} \right |}_{z=\pm h} = 0
\end{array}
\right.}
\Leftrightarrow
{\left\lbrace
\begin{array}{lcl}
{\left . \overline{\sigma}_{rz}^{H_1} \right |}_{z=\pm h} = 0, \\
{\left . \overline{\sigma}_{zz}^{H_0} \right |}_{z=\pm h} = 0.
\end{array}
\right.} \notag
\label{eq_BCElasticStress}
\end{equation}
Inserting Potential solutions Eq.~\eqref{eq_PotentialSolutions} into these equations provides the following linear equation for the constants $\mathcal{A}$, $\mathcal{B}$, $\mathcal{C}$, and $\mathcal{D}$:
\begin{widetext}
\begin{equation}
\mathcal{M} {\left(
\begin{array}{c}
\mathcal{A} \\
\mathcal{B} \\ 
\mathcal{C} \\
\mathcal{D}
\end{array}
\right)} =
{\left( 
\begin{array}{lcl}
-2 \chi \left(\frac{1}{\chi^2-p^2}\right) {\left (T_1e^{\chi h} - T_2e^{-\chi h} \right )} - 2\eta \left(\frac{\gamma}{1-\gamma^2\chi^2}\right)\left(\frac{\gamma^2}{1-\gamma^2p^2}\right )e^{-p h} e^{-h/\gamma} \\
-2  \chi \left(\frac{1}{\chi^2-p^2}\right) {\left (T_1e^{-\chi h} - T_2e^{\chi h} \right )} - 2 \left(\frac{\gamma}{1-\gamma^2\chi^2}\right)\left(\frac{\gamma^2}{1-\gamma^2p^2}\right)e^{-p h} e^{h/\gamma} \\ -(k^2 + q^2)e^{-p h} {\left[  \frac{1}{\chi^2 - p^2}{\left (T_1e^{\chi h} - T_2e^{-\chi h} \right )} - e^{-h/\gamma} \right]} \\ 
-(k^2 + q^2)e^{-p h} {\left[  \frac{1}{\chi^2 - p^2}{\left (T_1e^{-\chi h} - T_2e^{\chi h} \right )} - e^{h/\gamma} \right]}
\end{array}
\right)},
\label{eq_ConstantesABCD}
\end{equation}
where the constants $T_1$ and $T_2$ are given by Eq.~\eqref{eq_ConstantsT1T2} and
\begin{equation}
\mathcal{M} = {\left(
\begin{array}{cccc}
-2p e^{-2p h} & 2p & -(k^2 + q^2)e^{-(p + q)h} & -(k^2 + q^2)e^{-(p - q)h}\\
-2p & 2p e^{-2p h} & -(k^2 + q^2)e^{-(p - q)h} & -(k^2 + q^2)e^{-(p + q)h}\\
(k^2 + q^2)e^{-2p h} & (k^2 + q^2) & 2k^2q e^{-(p + q)h} & -2k^2q e^{-(p - q)h}\\
(k^2 + q^2) & (k^2 + q^2)e^{-2p h} & 2k^2q e^{-(p - q)h} & -2k^2q e^{-(p + q)h}
\end{array}
\right ).}
\label{eq_M}
\end{equation}
\end{widetext}
The determinant of the matrix $\mathcal{M}$ vanishes for Lamb waves, leading to singularities. To avoid this problem, a small damping factor is added to the angular frequency $\omega$. When not specified, it is taken equal to $6 \times 10^{-4}$ \dbus. Once the constants determined, the displacement amplitude at frequency $\omega$ is recovered by inverse Hankel transform Eq.~\eqref{eq_HankelFourierPotentialToDisplacementEquations}.
\begin{equation}
{\left\lbrace
\begin{array}{lcl}
\overline{u}_r(r,z,\omega) = \displaystyle \int_0^{\infty} \overline{u}_r^{H_1}(k,z,\omega)J_1(k r)k dk, \\
\overline{u}_z(r,z,\omega) = \displaystyle \int_0^{\infty} \overline{u}_z^{H_0}(k,z,\omega)J_0(k r)k dk.
\end{array}
\right. } \notag
\label{eq_displacement}
\end{equation}


\end{document}